\documentclass[twocolumn,showpacs,preprintnumbers,amsmath,amssymb]{revtex4}

\usepackage{graphicx}
\usepackage{dcolumn}
\usepackage{bm}
\usepackage{psfig}
\usepackage{epsfig}

\begin{document}

\newcommand{\vAi}{{\cal A}_{i_1\cdots i_n}}
\newcommand{\vAim}{{\cal A}_{i_1\cdots i_{n-1}}}
\newcommand{\vAbi}{\bar{\cal A}^{i_1\cdots i_n}}
\newcommand{\vAbim}{\bar{\cal A}^{i_1\cdots i_{n-1}}}
\newcommand{\htS}{\hat{S}}
\newcommand{\htR}{\hat{R}}
\newcommand{\htI}{\hat{I}}
\newcommand{\htB}{\hat{B}}
\newcommand{\htD}{\hat{D}}
\newcommand{\htV}{\hat{V}}
\newcommand{\cT}{{\cal T}}
\newcommand{\cM}{{\cal M}}
\newcommand{\cMs}{{\cal M}^*}
\newcommand{\vk}{{\vec k}}
\newcommand{\vK}{{\vec K}}
\newcommand{\vb}{{\vec b}}
\newcommand{{\vp}}{{\vec p}}
\newcommand{{\vq}}{{\vec q}}
\newcommand{\vQ}{{\vec Q}}
\newcommand{\vx}{{\vec x}}
\newcommand{\tr}{{{\rm Tr}}}
\newcommand{\beq}{\begin{equation}}
\newcommand{\eeq}[1]{\label{#1} \end{equation}}
\newcommand{\half}{{\textstyle \frac{1}{2} }}
\newcommand{\lton}{\mathrel{\lower.9ex \hbox{$\stackrel{\displaystyle
<}{\sim}$}}}
\newcommand{\gton}{\mathrel{\lower.9ex \hbox{$\stackrel{\displaystyle
>}{\sim}$}}}
\newcommand{\ee}{\end{equation}}
\newcommand{\ben}{\begin{enumerate}}
\newcommand{\een}{\end{enumerate}}
\newcommand{\bit}{\begin{itemize}}
\newcommand{\eit}{\end{itemize}}
\newcommand{\bc}{\begin{center}}
\newcommand{\ec}{\end{center}}
\newcommand{\bea}{\begin{eqnarray}}
\newcommand{\eea}{\end{eqnarray}}
\newcommand{\beqar}{\begin{eqnarray}}
\newcommand{\eeqar}[1]{\label{#1}\end{eqnarray}}
\newcommand{\bra}[1]{\langle {#1}|}
\newcommand{\ket}[1]{|{#1}\rangle}
\newcommand{\norm}[2]{\langle{#1}|{#2}\rangle}
\newcommand{\brac}[3]{\langle{#1}|{#2}|{#3}\rangle}
\newcommand{\hilb}{{\cal H}}
\newcommand{\pleft}{\stackrel{\leftarrow}{\partial}}
\newcommand{\pright}{\stackrel{\rightarrow}{\partial}}

\begin{flushright}
\vskip .5cm
\end{flushright} \vspace{1cm}

\title{Energy Systematics of Jet Tomography at RHIC: \protect{$\sqrt{s}=$} 62.4 vs 200 AGeV}

\author{A.~Adil}%
\email{azfar@phys.columbia.edu}

\author{M.~Gyulassy}
\email{gyulassy@phys.columbia.edu}

\affiliation{
Columbia University,
Department of Physics, 538 West 120-th Street, New York, NY 10027}%

\date{\today}

\begin{abstract}

The collision energy dependence of jet tomography is investigated
within the GLV formalism. The emphasis is on estimating
systematic uncertainties resulting from the interplay of energy
loss fluctuations and the rapid increase of the parton transverse
momentum slopes as  $\sqrt{s}$ decreases from 200 to 62 AGeV.
 \end{abstract}


\maketitle

\section{Introduction}

We study the energy dependence systematics of 
jet tomography in nuclear collisions
within the GLV formalism\cite{Gyulassy:2003mc,Gyulassy:2001nm}
in the $\sqrt{s}=$ 62 to 200 AGeV range.
In general, tomography provides a map of the density of matter through the
study of the angular pattern of attenuation of a calibrated source beam.
Extensive high $p_T$ data on single and dijet azimuthal correlations
and relative to the reaction plane have already been compiled
at 200 AGeV \cite{Adler:2003qi,Adams:2003kv} and used to determine
opacity and density of the quark-gluon plasma (QGP) produced in central Au+Au
reactions\cite{Vitev:2002pf}. The interest in applying
the same techniques to lower 62 AGeV data \cite{Back:2004ra} is
to test experimentally the predicted $\sqrt{s}$ decrease of the QGP density
and the predicted variation of the gluon/quark jet source.
In this paper, we calculate the nuclear modification factor,
$R_{AA}(p_{T}, y=0, \sqrt{s})$, for central $Au-Au$
collisions at both $\sqrt{s}=62.4,\; 200$ AGeV for neutral pions.
Previous predictions for 62 AGeV have been published by Wang\cite{Wang:2003aw}
and Vitev\cite{Vitev:2004gn}.

We concentrate here on the role of energy loss fluctuations
\cite{Gyulassy:2001nm,Wiedemann:2003} on the predicted single
hadron attenuation pattern in order to
gain an estimate of some of the systematic theoretical errors in the
jet tomographic technique.  To isolate
the role of fluctuations we neglect $k_T$ smearing, the Cronin
enhancement, gluon and quark shadowing, and nonperturbative baryon
dynamical contributions that strongly distort the hadron spectra below
$p_{T}< 4-5$ GeV\cite{Gyulassy:2003mc}. Therefore,
our calculations can only be compared to unidentified charged
particles  outside the baryon anomaly region\cite{Vitev:2001zn},
i.e. $ p_{T}\gton 5$ GeV. Even for the identified $\pi^0$ spectra,
the calculations neglecting the rapidly rising Cronin effect with
decreasing energy limits the applicability to the $p_{T}> 4$ GeV
region. See different estimates of those effects in
Refs.\cite{Wang:2003aw,Vitev:2004gn}. The intermediate $p_{T}$ is
also strongly influenced by the
  transverse Doppler shifted hydrodynamic
flow of the QGP bulk matter\cite{Hirano:2003yp}
as well as possible coalescence hadronization mechanisms\cite{Fries:2003vb}
not described by attenuated parton fragmentation.

The tomographic information about the maximum densities of the QGP
attained at RHIC as a function of energy
is best isolated from the $ p_T> 5$ GeV region.
However, it is important to examine critically
the underlying assumptions of jet tomography to assess
the theoretical systematic errors in its application to
nuclear reactions. We test below
the influence of the shape of the energy loss fraction
spectrum, $P(\epsilon, \bar{\epsilon})$
about the mean energy loss fraction, $\bar{\epsilon}$. The tomographic information
is encoded in $\bar{\epsilon}$ because
for Bjorken expansion it is is proportional to the
produced gluon rapidity density, $dN_g/dy$ as well as the path length.
Neglecting fluctuations as in \cite{Gyulassy:2000gk},
 i.e., $P(\epsilon, \bar{\epsilon})=
\delta(\epsilon- \bar{\epsilon})$, a fit of $R_{AA}$ provides a biased measure of
$\bar{\epsilon}$ and hence $dN_g/dy$.

As first emphasized in \cite{Baier:2001yt} fluctuations about
$\bar{\epsilon}$ must be considered in realistic applications because
the very rapid decrease of the transverse momentum spectra
of partons induces a bias to lower effective $\epsilon$. In Ref.\cite{Gyulassy:2001nm}
we quantified the magnitude of the fluctuation bias
in the GLV approach through a renormalization factor $Z$. We showed that for
130 AGeV conditions $Z\sim 0.5$ due to this bias, and hence the fitted $
\bar{\epsilon}$ from $R_{AA}$ must be upward corrected by a factor $1/Z \approx 2$.

At lower energies, the high $p_T$ slopes of the transverse momentum spectra
increase more and more as $\sqrt{s}$ decreases as seen in Fig. 1. 
Thus the calibrated gluon and quark source beams change systematically 
as a function of $\sqrt{s}$. This increases the fluctuation biases
and hence it is important
to investigate how sensitive is the tomographic analysis to
not only $\bar{\epsilon}$ but also to the {\em shape} of that distribution.

\section{Calculation of Spectra}

The neutral pion invariant inclusive cross section in $pp$ at
various values of $\sqrt{s}$ can be calculated via
conventional collinear factorized pQCD,

\begin{eqnarray}
\label{field} &&E_{h}\frac{d\sigma_{\pi^0}^{pp}}{d^3p} =K
        \sum 
        \int\!\!dx_1 dx_2  \;  
f_{a/A}(x_1,Q^2) f_{b/A}(x_2,Q^2)\ \nonumber \\
&\;& \hspace{6ex}            \frac{d\sigma^{ab \rightarrow
cd}}{d{\hat t}}
   \frac{D_{\pi^{0}/c}(z_{c},Q^2)}{\pi z_{c}} \,\,\, ,
\end{eqnarray}

The inclusive number distribution, $dN^{\pi^0}/dyd^2p_T$, in
$A+A$ collisions is obtained in the absence of nuclear effects by
simply multiplying Eq. \ref{field}  by the  Glauber geometric
(binary collision density) factor, $T_{AB}(b)$.  We use standard
Leading Order MRS D- parton distribution functions $f_{a}(x)$ (as
in \cite{MRSD:1998}), and KKP fragmentation functions
$D_{h}(z_{c})$ (as in \cite{KKP:1995}). Here, $x_{1}$ is the
momentum fraction of the projectile parton in the collision while
$x_{2}$ is the momentum fraction of the target parton.  The
$K(\sqrt{s})$ is a factor that simulates higher order effects.

\begin{figure}
\centering
 \epsfig{file=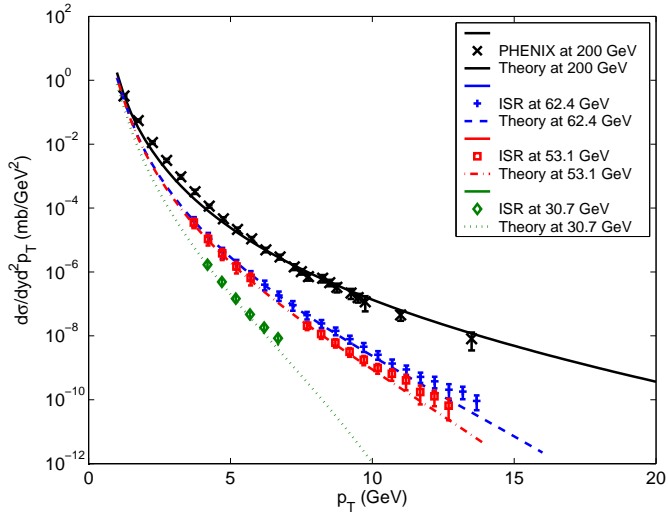,width=3.5in,angle=0}
  \caption{Invariant cross-sections for $\pi^{0}$ production in p-p collisions at $\sqrt{s}=30.7,53.1,62,4,200$ GeV.
  Data is from PHENIX\protect{\cite{Adler:2003qi}} (200 GeV) and ISR (all other energies) collaborations.}
  \label{unqpion}
\end{figure}

Figure \ref{unqpion} compares the invariant pQCD cross-sections for
$\pi^{0}$ production to available data.  The particular values used for $K$ and
$Q^{2}$ for the fits are given $K=2.0,Q^{2}=p_{T}^{2}$ for 200
GeV, $K=2.5,Q^{2}=0.5p_{T}^{2}$ for 62.4 GeV,
$K=3.5,Q^{2}=0.5p_{T}^{2}$ for 53.1 GeV and
$K=4.5,Q^{2}=0.5p_{T}^{2}$ for 30.7 GeV. The rapid increase of the slopes
with decreasing $\sqrt{s}$ is well reproduced by pQCD. However, the normalization
corrections due to NLO processes increases as $s$ decreases.

\section{Final State Quenching and Fluctuations}

In a dense QCD medium the induced radiative energy loss
reduces the initial transverse momentum $p_{T}^{0}$
of a produced hard parton  prior to hadron fragmentation.
The reduction factor $(1-\epsilon)$ depends in general on
both the initial parton rapidity $y$ and transverse momentum $p_{T}^0$
as well as on local QGP density
that varies with $s$ and impact parameter $b$ \cite{Gyulassy:2003mc}.

In this paper, we consider only mid-rapidity $(y=0)$ and central $b<<R$
collisions and further simplify the calculation by neglecting the slow $p_T$ dependence
of the energy loss fraction, $\epsilon$. The parton before fragmentation
has
$p_{T}^{f}={p_{T}^{0}}{(1-\epsilon)}$ in this picture. We neglect acoplanarity as well.

Fluctuation in the energy loss is taken into account
via a  probability
function $P(\epsilon,\bar{\epsilon})$ using \cite{Gyulassy:2001nm}
\begin{eqnarray}
\label{fullaa2} &&E_{h}\frac{d\sigma_{\pi^0}(\bar{\epsilon})}{d^3p} =K
        \sum 
        \int\!\!dx_1 dx_2  \;  
f_{a/A}(x_1,Q^2) f_{b/A}(x_2,Q^2)\ \nonumber \\
&\;& \hspace{6ex}            \frac{d\sigma^{ab \rightarrow
cd}}{d{\hat t}}
 \int d\epsilon \; P(\epsilon,\bar{\epsilon})
\frac{z^*_c}{z_c}
   \frac{D_{\pi^0/c}(z^*_c,Q^2)}{\pi z_c} \,\,\, ,
\end{eqnarray}
where $z_c^*=p_{T\pi}/p^f_{Tc}=z_c/(1-\epsilon)$. The parameter
$\overline{\epsilon}$ determines the magnitude of the quenching
and is expected decrease as the local density $\propto dN_g/dy$
decreases at lower energies. The $\overline{\epsilon}$ is
interpreted as the {\em average} $p_{T}$ shift.

In \cite{Gyulassy:2001nm} and \cite{Vitev:2004gn},
$P(\epsilon)$ was computed assuming Poisson fluctuations of the
radiated gluon number. This distribution is thus characterized
by the mean number of radiated gluons, $\langle N_g \rangle$,
as well as the mean energy loss $\bar{\epsilon}$. An inherent problem
of treating fluctuations via a Poisson process is that
it leads to violations of the energy bound $\epsilon =\sum_{n=1}^{N_g} \epsilon_n < 1$.
For small $\bar{\epsilon}_g<0.5 $ this is a small effect that can be
approximately corrected by truncating the distribution at $\epsilon=1$.
However, for high opacity QGP matter produced at RHIC, 
the gluon jet $\bar{\epsilon}_g\sim 0.6-0.8$ approaches unity\cite{Vitev:2004gn} 
and the truncation scheme leads to an approximately uniform distribution limiting 
 $\bar{\epsilon}_g<0.5$.

The truncated Poisson ansatz for fluctuations provides a natural way to
impose unitary and kinematic constraints on the induced radiative
energy loss. However, it is possible that other distributions could be operative in
the opaque limit. In this paper we explore two simplified forms of
fluctuation distributions to assess some of the the systematic uncertainties
in the predicted quenching factors. One is a simplified
``uniform'' model that reproduces the essential features of the truncated 
Poisson ansatz of \cite{Gyulassy:2001nm,Vitev:2004gn}. The second
distribution is called ``squeezed'' because it  accumulates
strength near $\epsilon\approx 1$ in the opaque limit.

The ``uniform'' distribution is assumed to have the form
\begin{eqnarray}
\label{unif} &&P(\epsilon,\overline{\epsilon})=\left\{
\begin{array}{ll}
\frac{\theta(0<\epsilon<2\bar{\epsilon})}{2\bar{\epsilon}}
& {\rm if }\;\; 0<\bar{\epsilon}<0.5
\nonumber \\
1 & {\rm if }\;\; 0.5<\bar{\epsilon}
\end{array} \right.
\end{eqnarray}
The ``squeezed'' distribution is assumed to have the form
\begin{eqnarray}
\label{squez} &&P(\epsilon,\overline{\epsilon})=\left\{
\begin{array}{ll}
\frac{\theta(0<\epsilon<2\bar{\epsilon})}{2\bar{\epsilon}} & {\rm
if }\;\; 0<\bar{\epsilon}<0.5
\nonumber \\
\frac{\theta(2\bar{\epsilon}-1<\epsilon<1)}{2(1-\bar{\epsilon})}
&{\rm if}\;\; 0.5<\bar{\epsilon}
\end{array} \right.
\label{squez}
\end{eqnarray}
Multiple gluon convolutions lead to phase space suppression
$\epsilon^n$ of low total energy loss, which is mimicked in (\ref{squez}) by
a simple $\theta$ function. A more detailed analysis of branching versus truncated
Poisson distributions will be presented elsewhere\cite{AG:2004}. For our present 
purposes of gaining a first measure of systematic errors the above overly simplified
forms are adequate. Note that while
the uniform distribution has a limiting $\bar{\epsilon}\le \half$ even for gluons,
 the ``squeezed'' one can easily accomodate the more realistic $\bar{\epsilon}_g>0.5$
situation at RHIC.

The shape of $P(\epsilon,\bar{\epsilon})$ 
depends not only on $\bar{\epsilon}$ but also on $<N_{g}>$, 
the average number of total gluons
emitted.  A smaller value of $<N_{g}>$
weights the lower $n$ values more and leads to a more uniform
$P(\epsilon,\bar{\epsilon})$ while larger values of $<N_{g}>$
weight higher $n$ values \cite{AG:2004}.  The average number of
gluons radiated increased with $\sqrt{s}$. Thus, we expect the
uniform distribution to be a better approximation for
$P(\epsilon,\bar{\epsilon})$ at 62.4 GeV while the ''squeezed''
distribution could be more realistic at 200 GeV.

The nuclear modification factor can then be calculated
for either distribution from Eq.
\ref{fullaa2} via
\begin{equation}
R_{AA}(p_T, s) =
\frac{d\sigma_{\pi^0}(\bar{\epsilon}(s))}{d\sigma_{\pi^0}(0)}
\label{raamine}
\end{equation}

\subsection{\protect{$R_{AA}$} with Squeezed Fluctuations}
\begin{figure}
  \centering
\epsfig{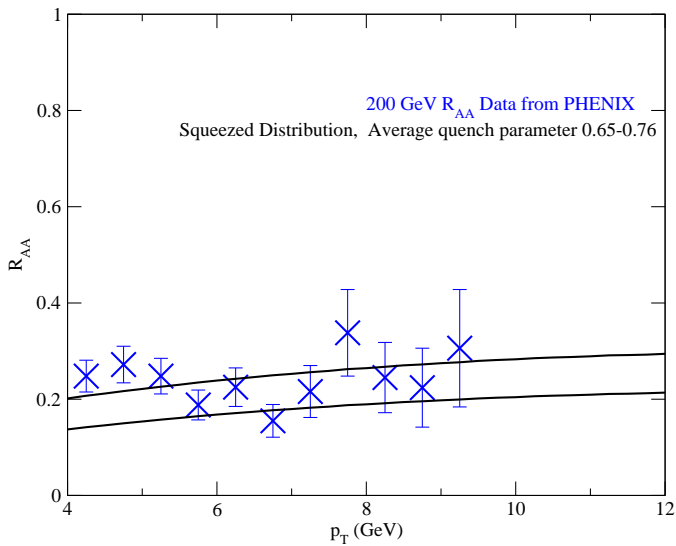}
  \caption{$R_{AA}(p_T,\sqrt{s}=200)$ with squeezed fluctuations
\protect{\ref{squez}}. The $\pi^0$ data are from PHENIX\protect{\cite{Adler:2003qi}}}
  \label{RAAbandfluc}
\end{figure}

Our first step is to fit $\bar{\epsilon}(200)$ to the 200 AGeV PHENIX data\cite{Adler:2003qi}
in the region $p_T\sim 8$ GeV. Figure \ref{RAAbandfluc} shows the
resulting $R_{AA}$.  The quenched pQCD prediction for
lower $p_{T}< 6$ GeV increasingly deviate from the data
 because intrinsic   $k_{T}$ smearing and
Cronin calculations are neglected here. However, the approximate
$p_T$ independence of higher $p_T$ is well accounted for as in
\cite{Vitev:2002pf}.  The band corresponds to an assumed average mean energy loss
fraction for gluons at 200 AGeV $0.65<\bar{\epsilon}_g(200)<0.76$ using the
squeezed distribution. Further below we will contrast the results to the energy evolution
with the uniform distribution.

The beam energy dependence of $\bar{\epsilon}(s)$ is predicted
to follow the energy dependence of the produced $dN_g/dy$.
We also take into account the Casimir difference between quark and gluon energy loss.
The mean energy loss fraction of parton type $c=q,g$ evolves with energy as
\begin{equation}
\bar{\epsilon}_c(\sqrt{s})=\frac{C_c}{C_g}\left(
\frac{{dN_{g}(\sqrt{s})}/{dy}}{{dN_{g}(200)}/{dy}}\right) \bar{\epsilon}_g(200)
\label{qparam}
\end{equation}
Thus $\bar{\epsilon}_q=4/9 \bar{\epsilon}_g$

The predicted $R_{AA}$ at 62 AGeV is obtained using the  multiplicity
systematics from PHOBOS\cite{Back:2001ae}
 that suggests a value for $dN_{tot}{dy}\approx$ 650-770.
From entropy
conservation we expect that the gluon rapidity density is comparably decrease from about 1000 at 200 AGeV. Therefore,  we expect that
$\bar{\epsilon}_g(62.4)\approx 0.65-0.77 \; \bar{\epsilon}_g(200)$.
\begin{figure}
  \centering
\epsfig{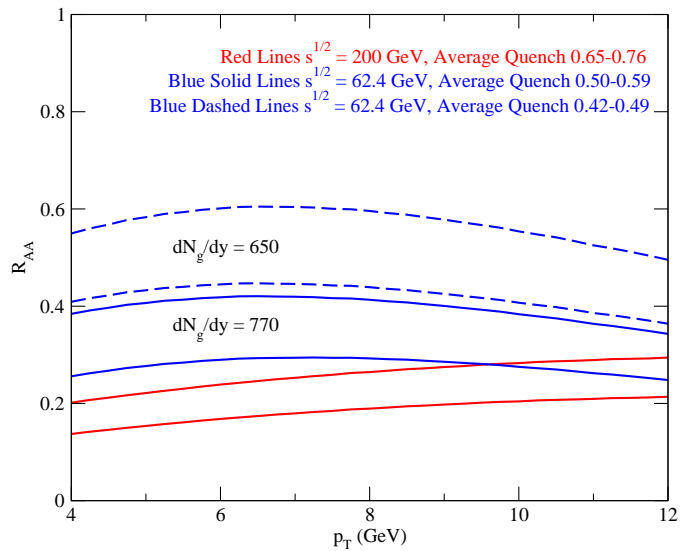}
  \caption{Predicted energy evolution of  $R_{AA}$ with
squeezed fluctuations at $\sqrt{s}=62.4,200$ GeV.}
  \label{RAAfluc}
\end{figure}
The bands of $R_{AA}$ predictions at 62.4 GeV are shown in Figure
\ref{RAAfluc}.  The ranges of average quenching shown in the
figure are $0.42<\bar{\epsilon}(62.4)<0.49$ and
$0.50<\bar{\epsilon}(62.4)<0.59$ for multiplicity values of 650
and 770 respectively.  These ranges are calculated by realizing
that $\frac{dN_{g}}{dy}(200)\approx1000$.

\subsection{\protect{$R_{AA}$} with Uniform Fluctuations}

We now repeat the $R_{AA}$ calculations in the previous section
using the uniform  distribution for
$P(\epsilon,\bar{\epsilon})$ from Eq.(\ref{unif}).
This is done to investigate the
sensitivity of the results to the form of this distribution.
 The uniform distribution is the same as  the
squeezed distribution for $\bar{\epsilon}_c<0.5$. Therefore the
quarks energy loss fraction distribution is identical for both
distributions. However, since the gluon GLV energy loss fraction
$\bar{\epsilon}_g>0.5$ in this energy range, the uniform
distribution actually saturates the mean energy loss fraction of
gluons at $\bar{\epsilon}_g=1/2$. This contrasts with  the
squeezed distribution that squeezes the gluon fraction energy
loss toward the $\epsilon=1$ boundary.
 Therefore the gluon
quenching is reduced with the uniform distribution
while preserving the quark quenching. As we note further below,
this also changes the glue to quark contribution ratios at high $p_T$.
\begin{figure}
  \centering
\epsfig{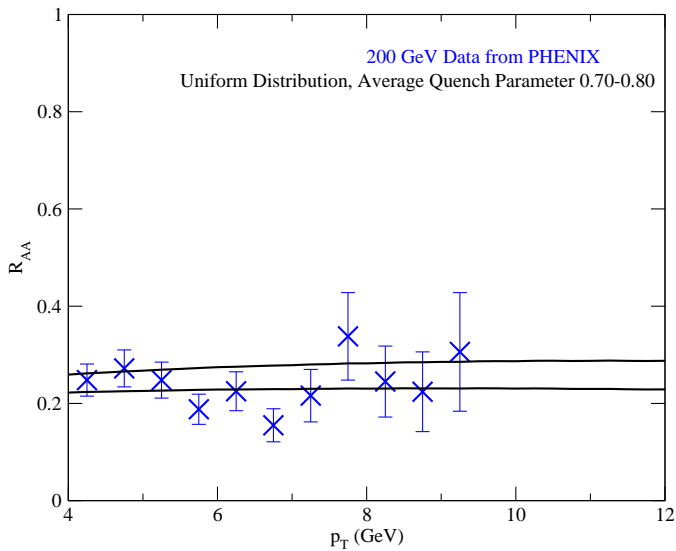}
  \caption{$R_{AA}$ at 200 GeV calculated using the uniform distribution.
The $\pi^0$ data are from PHENIX\protect{\cite{Adler:2003qi}}.}
  \label{RAAbandflucnew}
\end{figure}

$R_{AA}$ is compared  to PHENIX data using the uniform fluctuation distribution
in Figure \ref{RAAbandflucnew}.
The curves are distinctly flatter than the ones in the previous
section (Figure \ref{RAAbandfluc}).  This is because the
uniform quenching distribution does not suppress so much the gluon jets
than the squeezed distribution.
The band in this case corresponds to
$0.70<\bar{\epsilon}_g(200)<0.80$. In the uniform case
this implies that the mean quark energy loss fraction
at $\sqrt{s}=200$ is $0.31<\bar{\epsilon}_q(200)<0.36$
\begin{figure}
  \centering
\epsfig{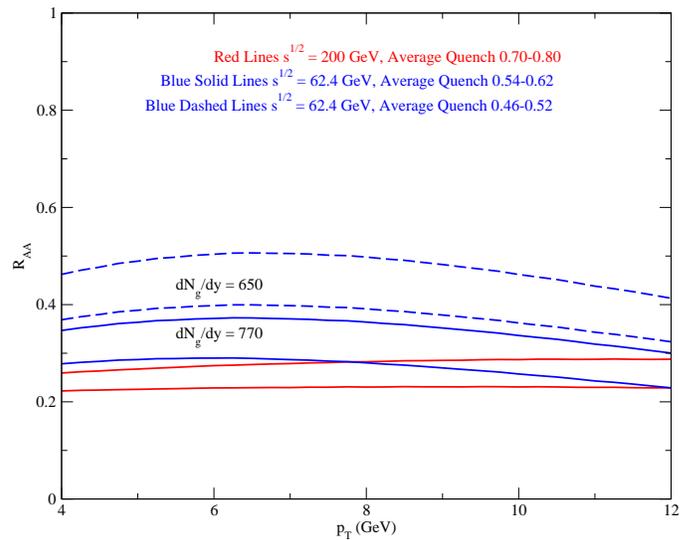}
  \caption{$R_{AA}$ at $\sqrt{s}=62.4,200$ GeV calculated using the uniform distribution.}
  \label{RAAflucnew}
\end{figure}

Figure \ref{RAAflucnew} shows the $R_{AA}$ prediction for 62 AGeV in this case.
 The range of mean gluon energy fractional energy loss is
$0.46<\bar{\epsilon}_g(62)<0.52$ and
$0.54<\bar{\epsilon}_g(62)<0.62$ for the  different estimates of
the $ dN_g/dy(62) =650 , 770$ at 62 AGeV respectively. The
nuclear modification factor has a weaker energy dependence in this
case and is generally somewhat smaller relative to the squeezed
fluctuation case discussed in the previous section. This is due
to the approximate saturation property of $\bar{\epsilon}_g \le
1/2$, as in the full Poisson fluctuation case \cite{Vitev:2004gn}.

These calculations for the nuclear modification factor at
$\sqrt{s}=62.4$ GeV are consistent with the calculations
performed in \cite{Vitev:2004gn}.  In the $p_{T}$ region 4-6 GeV
the prediction bands of both the squeezed and uniform
distributions are within the approximate range as predicted by
Vitev in \cite{Vitev:2004gn}.  The main differences in the
predictions can be explained by noting that the calculation in
\cite{Vitev:2004gn} includes models for Cronin interactions as
well as initial parton $k_{T}$ smearing which are not included in
the current calculation.  Both of these effects tend to deplete
hadrons in the higher transverse momentum regions and increase
their number in the intermediate transverse momentum region from
2-4 GeV.  This will explain the lower quenching at lower $p_{T}$
predicted by Vitev as compared to the current calculation.

Further comparisons can be attempted with recent data from the
PHOBOS collaboration as seen in \cite{Back:2004ra}.  This
publication, however, provides data at lower levels of $p_{T}$
and measures unidentified charged hadron cross-sections. This
leads to a double fold problem with the comparison.  Firstly, our
calculation does not apply in the kinematic region probed by
PHOBOS in \cite{Back:2004ra}.  Lower values of $p_{T}$ are not
modeled well by our calculation due to previously mentioned
reasons of neglecting Cronin effect and $k_{T}$ smearing. Also,
the calculation of charged hadrons would only be similar to
neutral pions where there were not many baryons present.  The
kinematic region probed by PHOBOS is dominated by baryonic data
which is not taken into account in the calculation.  This effect
would increase $R_{AA}$ as can be seen from the data shown in
\cite{Back:2004ra}.

\section{Comparison of $R_{AA}$ for different values of $\sqrt{s}$}

\begin{figure}[b]
  \centering
\epsfig{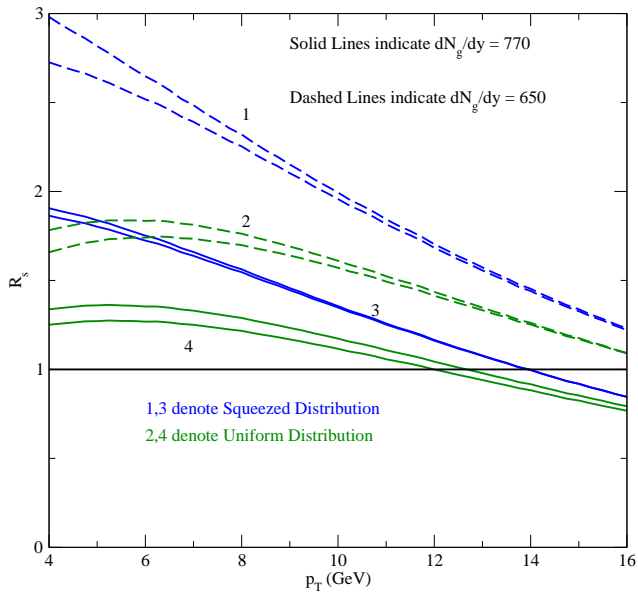}
  \caption{The ratio of $R_{AA}$ at 62.4 GeV to the one at 200 GeV
for different multiplicity assumptions. Curves for squeezed
distributions tend to be higher than curves for uniform
distributions.}
  \label{R-S}
\end{figure}

Note that  the predicted $R_{AA}(p_T,\sqrt{s}=62)$
 have a negative $p_T$ slope compared to the generally
flat $R_{AA}$ at 200 AGeV.  At very high $p_T$,
the 62.4 GeV curves even dip below the ones at 200 GeV.
This higher slope can be more robustly seen by calculating the variable
$R_s(s)$ as
\begin{equation}
R_{s}(s)= \frac{R_{AA}(s)}{R_{AA}(200)} \label{R-Seq}
\end{equation}
 showed in Figure \ref{R-S}.  The figure shows
multiple curves for both types of quench distributions.  One can
see that the $R_{s}$ curves have a distinct downward slope due to
the increasing power law index of the underlying pQCD parton
distributions at lower energies.
These indices vary much more with $p_{T}$ at 62
AGeV than at 200 AGeV. Therefore, the downward slope with
increasing $p_{T}$ is due to the increased slope of the input
spectra with increasing $p_{T}$ at lower $\sqrt{s}$.

It is
remarkable that even with a smaller average energy loss at 62 AGeV, the
more rapid decrease of high $p_{T}$ parton production
compensates to make it appear that there is 
greater quenching. The $\sqrt{s}$ systematic variation of the calibrated initial 
jet source transverse spectra used leads to this anti-intuitive
feature and must be carefully taken into account
in future tomographic inversion of single and di-jet quaenching.
Note that   $R_s$ much less sensitive to the propagation
of the uncertainty  in $R_{AA}$ at 200 AGeV because absolute $dN_g/dy$
uncertainty tends to cancel out in this $R_s$ ratio.
 This variable is also seen to be able to
differentiate clearly between these two types of fluctuation
 distributions.

\section{Quark vs Gluon Fragmentation}

An interesting theoretical measure of the underlying dynamics  is
shown in fig.(\ref{pig200},\ref{pig62})
\begin{figure}
  \centering
\epsfig{file=part_raa_200.eps,width=3.3in,angle=0}
  \caption{The gluon and quark contributions to the total
$R^{\pi^0}_{AA}(p_T,\sqrt{s}=200)$ are shown using a uniform distribution.
Gluon and quark fragmentation
gives comparable contributions to the $\pi^0$ yield at this energy and kinematic range.
}
  \label{pig200}
\end{figure}

\begin{figure}
  \centering
\epsfig{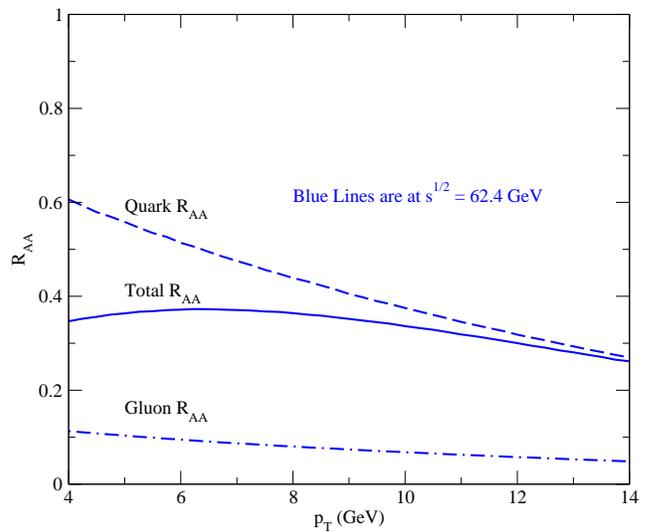}
  \caption{The gluon and quark contributions to the total
$R^{\pi^0}_{AA}(p_T,\sqrt{s}=62)$ are shown. In contrast to $\sqrt{s}=200$ AGeV
the $\pi^0$ at 62 AGeV are dominated by quark fragmentation.}
  \label{pig62}
\end{figure}

We see that while quenched quark and gluon fragmentation
contributes comparable amounts to the final quenched $\pi^0$ spectrum at 200 AGeV,
The gluon contribution is almost completely quenched by $62$ AGeV
in our calculations in spite of the smaller $\bar{\epsilon}_g$ at 62.
The quark and gluon $p_T$ transverse evolve with $s$ in different calculable ways in pQCD.
This evolution however causes the rapid disappearance of the gluonic contribution
as $s$ decreases in the case of the uniform distribution.

\begin{figure}
  \centering
\epsfig{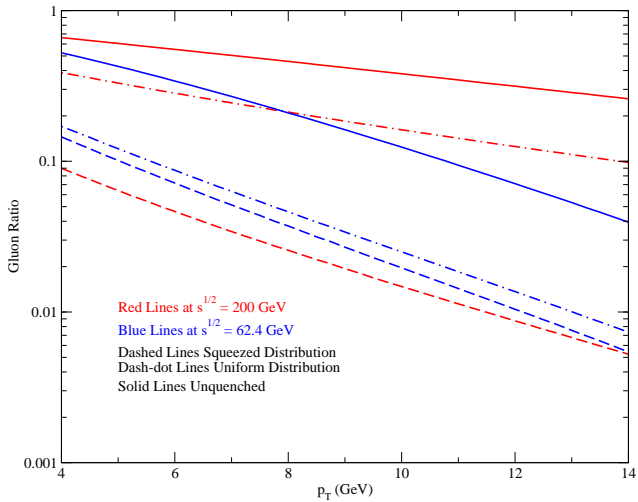}
  \caption{The ratio of $\pi^{0}$ from gluons both quenched and unquenched at $\sqrt{s}=62.4,200$ GeV.
  $\frac{dN_{g}}{dy}(62.4)$ is assumed to 770.  Lower levels of quenching are used to make this estimate.
  $\bar{\epsilon}_{\textrm{squeezed}}$ is 0.65 and 0.50 at 200 GeV and 62.4 GeV respectively. $\bar{\epsilon}_{\textrm{uniform}}$
   is 0.7 and 0.54 at 200 GeV and 62.4 GeV respectively.}
  \label{GluonRatios}
\end{figure}

In Fig.\ref{GluonRatios} we see that the form of the
fluctuation distribution strongly influences the pion from  gluon jet fraction.
For the squeezed distribution, the gluon jet contribution
is already negligible at 200 AGeV and of course remains so at 62
modulo a small increase due to the decrease of $\bar{\epsilon}(62)$ relative to 200.

These results suggest
that identified particle ratios sensitive to quark vs gluon fragmentation\cite{Fai:2001vz}
in the high $p_T>5$ GeV domain would provide valuable  to constrain
on the form of the energy loss fluctuations and to map out
more accurately the interplay of quark and gluon contributions.

\section{Conclusions}

Predictions for $R_{AA}(p_T,\sqrt{s}=62)$ for $\pi^0$ were made
with a focus on the systematic uncertainties resulting from as
yet un-tested assumptions about the energy loss fluctuation
spectrum. The most unintuitive aspect of the prediction is a
decreasing $R_{AA}$ with $p_T$ at lower energies that is due to
the more rapid decrease of the produced parton $p_T$
distributions. The negative $p_T$ slope of $R_{AA}$ provides an
important constraint on the fluctuation distributions. The pion
fractions from gluon jets were also shown to be sensitive to the
underlying fluctuation distributions. Identified particle ratios
at $p_T>5$ GeV should help to further test the distribution. Such
further studies are essential in order to calibrate more
accurately the jet tomographic measure of the initial QGP
densities produced at RHIC.

\begin{acknowledgments}
Discussions with I. Vitev and X.N. Wang are gratefully acknowledged.
  This work is supported in part by the United States
Department of Energy
under Grants   No. DE-FG02-93ER40764.
\end{acknowledgments}

\end{document}